\documentclass{cupconf}
\usepackage{graphicx}
\usepackage{subfigure}
\usepackage{natbib}

  \checkfont{eurm10}
  \iffontfound
    \IfFileExists{upmath.sty}
      {\typeout{^^JFound AMS Euler Roman fonts on the system,
                   using the 'upmath' package.^^J}%
       \usepackage{upmath}}
      {\typeout{^^JFound AMS Euler Roman fonts on the system, but you
                   dont seem to have the}%
       \typeout{'upmath' package installed. cupconf.cls can take advantage
                 of these fonts,^^Jif you use 'upmath' package.^^J}%
      }
  \else
  \fi

  \checkfont{msam10}
  \iffontfound
    \IfFileExists{amssymb.sty}
      {\typeout{^^JFound AMS Symbol fonts on the system, using the
                'amssymb' package.^^J}%
       \usepackage{amssymb}%

      }{}
  \fi

  \IfFileExists{amsbsy.sty}
    {\typeout{^^JFound the 'amsbsy' package on the system, using it.^^J}%
     \usepackage{amsbsy}}
    {}

\def\km/s{km~s$^{-1}$}

\def\Msun{\hbox{\it M$_\odot$}}
\def\Lsun{\hbox{\it L$_\odot$}}
\def\Myr{Myr}
\def\Gyr{Gyr}
\def\Minit{\hbox{M$_{\rm initial}$}}

\def\FMM362{FMM362}
\def\mnras{{\it Monthly Notices of the Royal Astronomical Society\, }}
\def\qjras{{\it Quarterly Journal of the Royal Astronomical Society\, }}
\def\aap{{\it Astronomy \& Astrophysics\, }}
\def\nat{{\it Nature\, }}
\def\apj{{\it Astrophysical Journal\, }}
\def\apjs{{\it Astrophysical Journal Supplements\, }}
\def\aj{{\it Astronomical Journal\, }}
\def\pasp{{\it Publications of the Astronomical Society of the Pacific\, }}
\def\apjl{{\it Astrophysical Journal Letters\, }}
\def\araa{{\it Annual Review of Astronomy and Astrophysics\, }}
\def\simgr{\mathrel{\hbox{\rlap{\hbox{\lower4pt\hbox{$\sim$}}}\hbox{$>$}}}}
\def\simls{\mathrel{\hbox{\rlap{\hbox{\lower4pt\hbox{$\sim$}}}\hbox{$<$}}}}
\def\micron{$\mu$m}
\long\def\symbolfootnote[#1]#2{\begingroup%
\def\thefootnote{\fnsymbol{footnote}}\footnote[#1]{#2}\endgroup} 

\newsavebox{\astrutbox}
\sbox{\astrutbox}{\rule[-5pt]{0pt}{20pt}}

\title[Massive Star Formation in the Galactic Center]{Massive Star Formation \\ in the Galactic Center}

\author[D. F. Figer {\it et al.\/}]%
{D\ls o\ls n\ls~F.\ns F\ls i\ls g\ls e\ls r%
}

\affiliation{Rochester Institute of Technology, Rochester, NY, USA\\[\affilskip]
}

\pubyear{2006}
\volume{1}
\pagerange{1}
\begin{document}

\maketitle

\begin{abstract}
The Galactic center is a hotbed of star formation activity, containing the most massive
star formation site and three of the most massive young star
clusters in the Galaxy. Given such a rich environment, it contains more stars with initial masses
above 100~\Msun\ than anywhere else in the Galaxy. This review concerns 
the young stellar population in the Galactic center, as it relates to 
massive star formation in the region. The sample includes stars in the three massive stellar clusters, the 
population of younger stars in the present sites of star formation, the stars surrounding the
central black hole, and the bulk of the stars in the field population. The fossil record
in the Galactic center suggests that the recently formed massive stars there are 
present-day examples of similar populations that must have been formed through 
star formation episodes stretching back to the time period when the Galaxy was forming. 
\end{abstract}

\firstsection 
\section{Introduction}
The Galactic center (GC) is an exceptional region for testing massive star formation and evolution models. 
It contains 10\% of the present star formation activity in the Galaxy, yet fills only
a tiny fraction of a percent of the volume in the Galactic disk\footnote{For the purposes of this review, the GC refers to 
a cylindrical volume with radius of $\approx$500~pc and thickness of
$\approx$60~pc that is centered on the Galactic nucleus and is coincident with a region of increased dust and 
gas density, often referred to as the ``Central Molecular Zone'' \citep{ser96}.}. The initial conditions for star
formation in the GC are unique in the Galaxy. The molecular clouds in the region are extraordinarly dense, under high
thermal pressure, and are subject to a strong gravitational tidal field. \citet{mor93} argue that these conditions may favor
the preferential formation of high mass stars. Being the closest galactic nucleus, the GC gives us
an opportunity to observe processes that potentially have wide applicability in other galaxies, both
in their centers and in the interaction regions of merging galaxies. Finally, the GC may be the
richest site of certain exotic processes and objects in the Galaxy, i.e.\ runaway stellar
mergers leading to intermediate mass black holes and stellar rejuvination through atmospheric stripping, to name
a few. 

This review is primarily concerned with massive star formation in the region. 
For thorough reviews on a variety of topics concerning the Galactic center, 
see \citet{gen87}, \citet{gen94}, \citet{mor96}, and \citet{eck05}.

\section{The Galactic center environment and star formation}
The star formation efficiency in the GC appears to be high. Plotting the surface star formation rate ($\Sigma_{\rm SFR}\sim$5~\Msun~yr$^{-1}$~pc$^{-2}$) versus 
surface gas density ($\Sigma_{H_2}\sim$400~\Msun~pc$^{-2}$) in a ``Schmidt plot'' suggests an efficiency of nearly 100\%, comparable to that of
the most intense infrared circumnuclear starbursts in other galaxies and a factor of
twenty higher than in typical galaxies \citep[see Figure~7 in][]{ken98}. It is also higher
than that elsewhere in the Galaxy; commensurately, stars in the GC emit about 5-10\% of the 
Galaxy's ionizing radiation and infrared luminosity. 

\citet{mor96} review the content and conditions of the interstellar medium in the ``Central Molecular Zone'' (CMZ), 
noting that the molecular clouds in the region are extraordinarily dense ($n>10^4~cm^{-3}$) and
warm ($T\sim70~K$) with respect to those found in the disk of the Galaxy. \citet{sta89} argue
that the density and internal velocities of clouds in the GC are a direct
result of the strong tidal fields in the region, i.e.\ only the dense survive. \citet{ser96} argue that the
inexorable inflow of molecular material from further out in the Galaxy powers 
continuous and robust star formation activity in the region. 

It is still unclear how magnetic field strength affects star formation. If it does matter, then the
GC might be expected to reveal such effects. 
The strength of the magnetic field in the GC has been estimated through far infrared polarized light
from aligned dust grains \citep{hil93,chu03} and Zeeman splitting of the OH molecule \citep{pla95}.
In both cases, the field is inferred to be of milliGuass strength. However, \citet{uch95} argue
strongly that these strengths are localized
to bundles that delineate the extraordinary non-thermal filaments in the region \citep{yus87}, and are
not representative of the field strength that is pervasive in the region. If this is correct, then
the fields inside GC molecular clouds may not be so strong versus those inside disk clouds ($B\sim3~\mu$G).

Metals in molecular clouds can provide cooling that aids protostellar collapse, but they also
create opacity to the UV flux, winds, and bipolar outflows that emanate from newly formed stars. 
Measurements of metallicty in the Galactic center span a range of solar, observed
in stars \citep{ram00,car00,naj04}, to twice solar, observed in the gas phase \citep{shi94},
to four times solar, observed through x-ray emission near the very center \citep{mae02}.
The errors from the stellar measurements are the smallest and suggest that stars in the 
GC are formed from material with roughly solar abundances. 

\section{Present-day star formation in the GC}
\begin{figure}
\begin{center}
\vskip 0.7truein
\includegraphics[height=10cm]{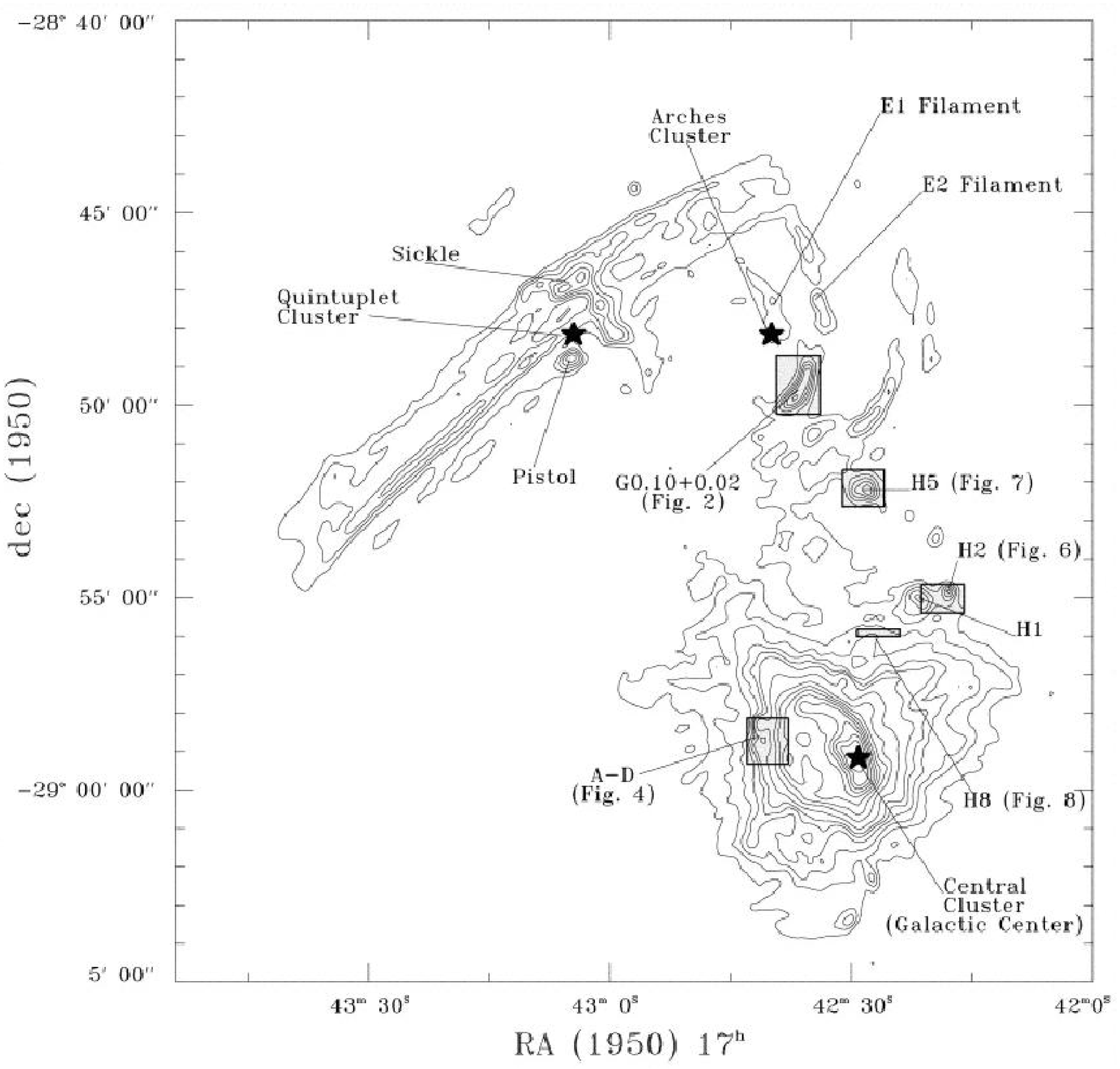}
\caption{Radio emission from the GC region at 6 cm, adapted in Figure~1 by \citet{cot99} from \citet{yus87}. The star
symbols represent the three massive clusters. Hot stars in the Quintuplet and Arches
clusters ionize gas on the surfaces of nearby molecular clouds to produce the radio emission in the ``Sickle'' and ``G0.10+0.02/E1/E2 Filaments,''
respectively. The radio emission near the Galactic center is due to a combination
of thermal and non-thermal emission. The ``H1-8'' and ``A-D'' regions are ultra-compact HII regions
surrounding recently formed stars.\label{radio}}
\end{center}
\end{figure}

Present-day star formation in the GC is somewhat subdued compared to the 
episodes that produced the massive clusters we now see. A dozen or so ultra-compact
HII regions are distributed throughout the central 50~pc, each containing one or a few O-stars
still embedded in their natal environs. \citet{yus87} identify most of these sources in radio
continuum observations (see Figure~\ref{radio}). \citet{zha93} and \citet{gos85} infer lyman-continuum fluxes that are comparable to
that expected from a single O7V star in each of the H1-H5 and A-D UCHII regions. \citet{cot99} find
that several of the recently formed stars in these regions have broken out of their dust
shroud, revealing spectra of young massive stars; see also \citet{fig94} and \citet{mun06} for additional examples.

\begin{figure}
\begin{center}
\vskip 0.3truein
\includegraphics[width=7cm]{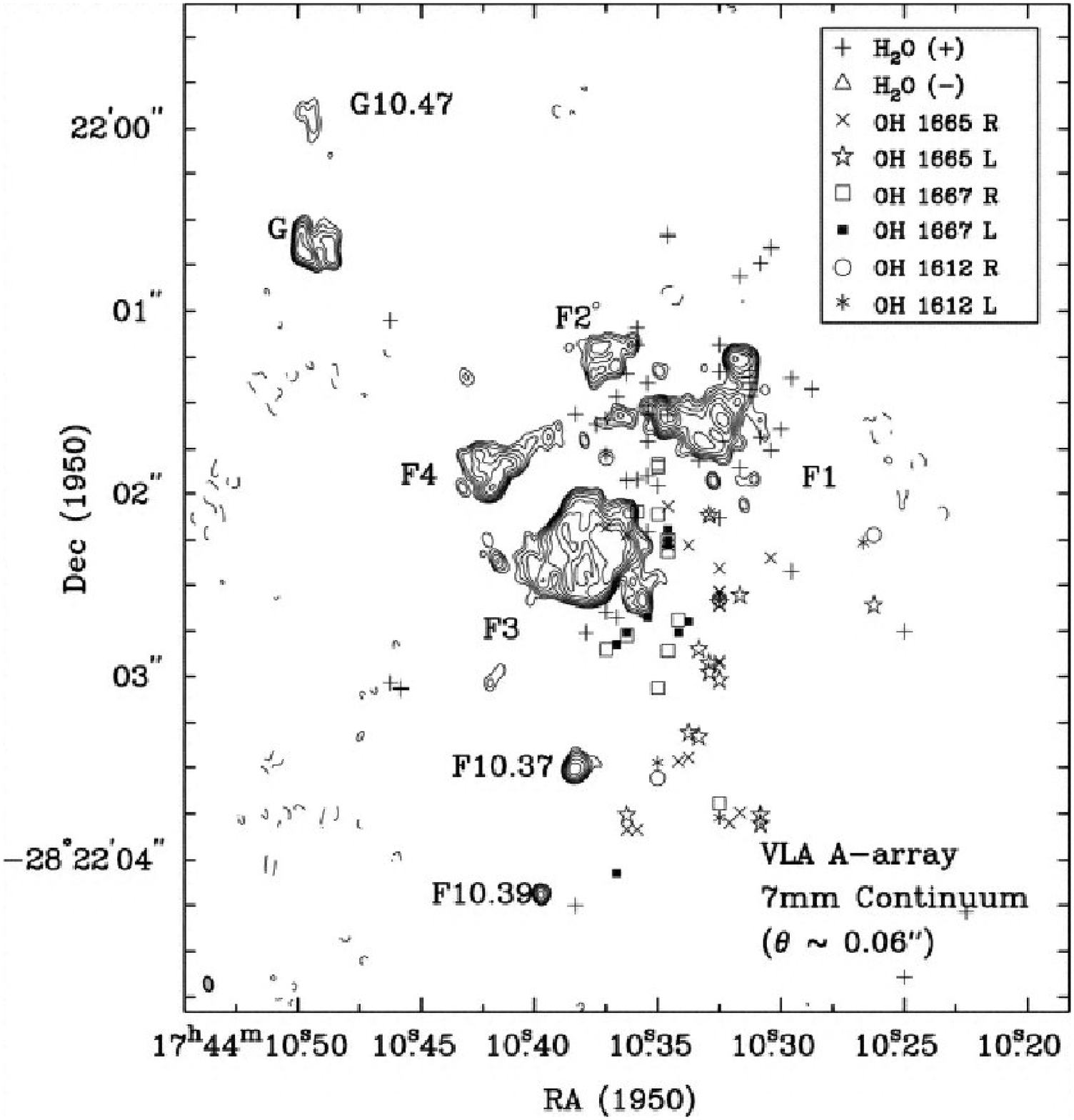}
\caption{Figure~3 from \citet{mcg04} showing H$_2$O and OH masers overplotted on 7~mm contours for a 
small portion of the Sgr~B2 cloud. The activity in this region is typical of that found near the fifty or
so ultra-compact HII regions in Sgr~B2.
\label{mcgrath}}
\end{center}
\end{figure}

A bit further from the GC, the Sgr~B2 molecular cloud harbors a massive star cluster in the making and is home 
to the most intense present-day star formation site in the Galaxy \citep{gau95,dep95,mcg04,tak02,dev00,liu99,gar99,dep96}. 
Within the next few Myr, this activity should produce a
star cluster that is comparable in mass to the Arches cluster (see Figure~\ref{mcgrath}). 
\citet{sat00} note evidence in support of a cloud-cloud collision as the origin for the
intense star formation in Sgr~B2; these include velocity gradients, magnetic field morphology,
shock-enhanced molecular emission, shock-induced molecular evaporation from dust grains, and
distinctly different densities of certain molecular species throughout the cloud.

\section{Continuous star formation in the GC}
\begin{figure}
\begin{center}
\vskip 0.3truein
\includegraphics[width=6cm]{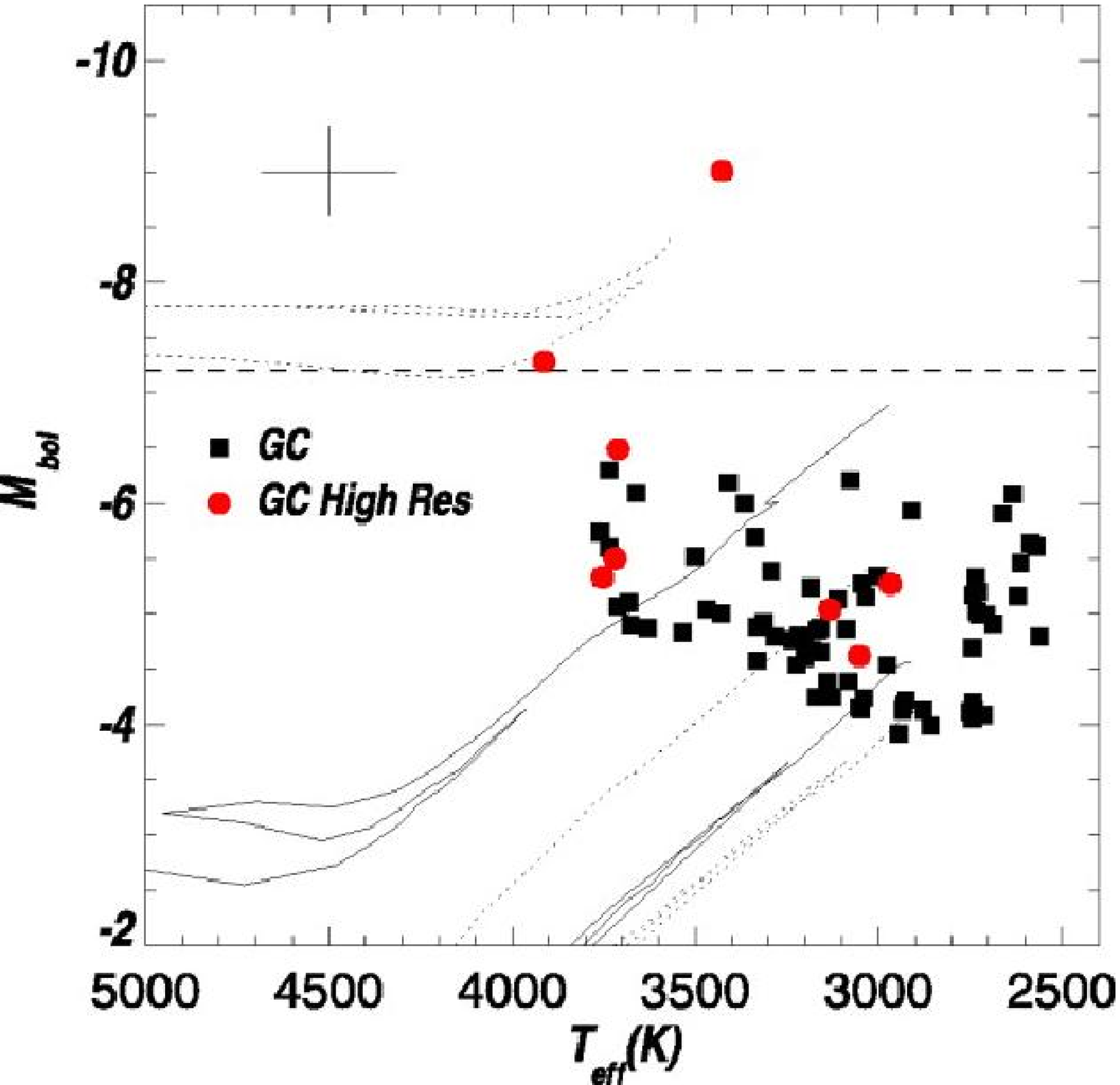}
\caption{Estimates of absolute magnitude versus temperature for stars in the
GC from \citet{blu03}.
The lines correspond to model isochrones having ages of 10~\Myr, 100~\Myr, 1~\Gyr, 5~\Gyr,
and 12~\Gyr. The supergiants (above the horizontal line) are descendant from stars
having M$\approx$15-25~\Msun, whereas fainter stars are descendant from lower
mass main sequence stars having a few to 15~\Msun. The presence of these stars in the GC demonstrates
intermediate age star formation of massive stars.\label{blum1}}
\end{center}
\end{figure}

There is ample evidence for persistent star formation in the GC in the form of upper-tip asymptotic giant
branch stars distributed throughout the region \citep{leb87,nar96,fro99,sjo99}. Figure~\ref{blum1} shows a plot for some of
these stars, based on spectroscopic data, overlaid with intermediate age model isochrones \citep{blu03}. 
Note that the giants and supergiants in
this plot require ages that span a few \Myr\ to a few \Gyr.

\begin{figure}
\begin{center}
\vskip 0.3truein
\includegraphics[width=10cm]{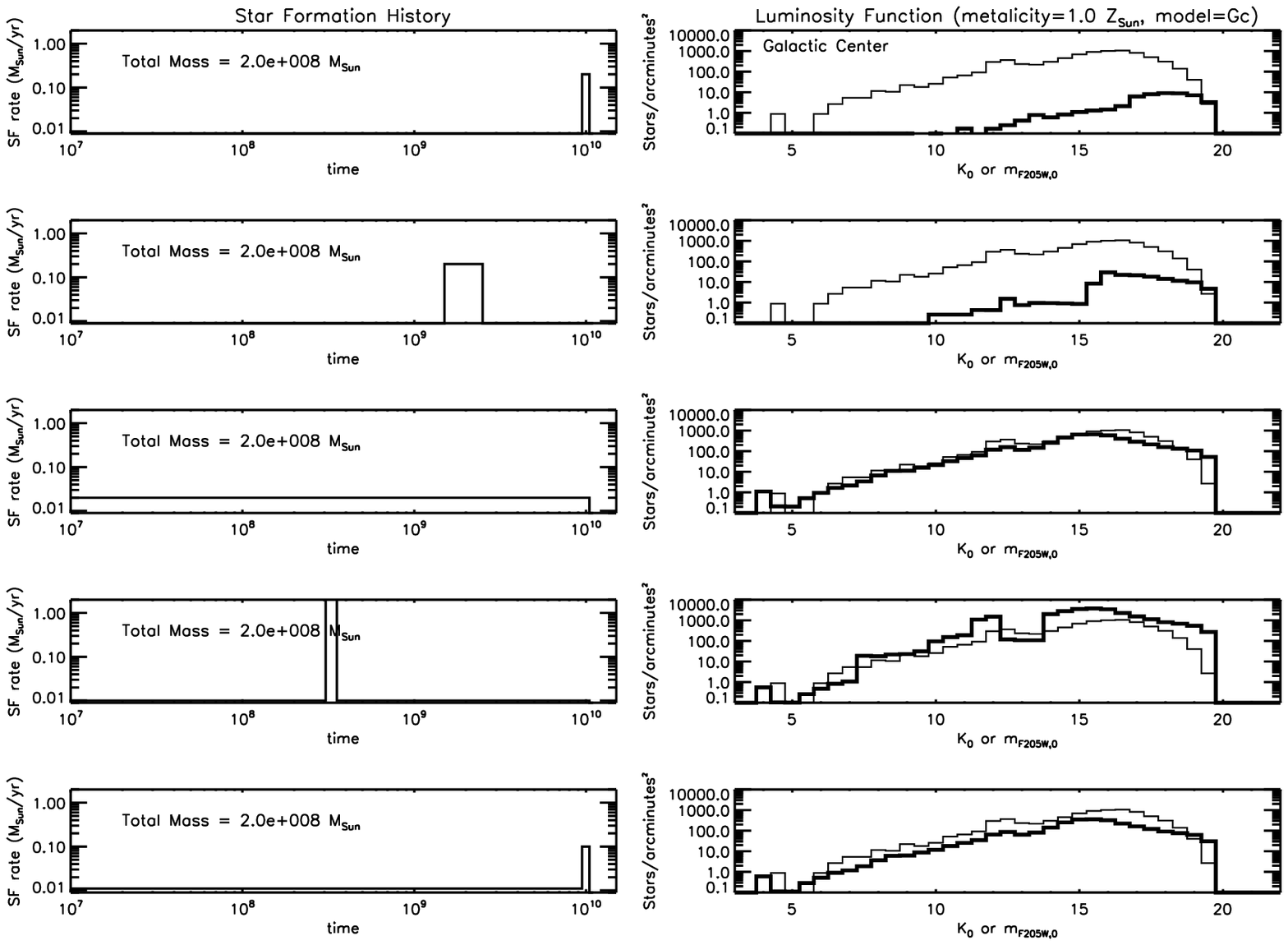}
\caption{A figure adapted from \citet{fig04b} showing various star formation scenarios ({\it left}), and resultant model luminosity functions ({\it right, thick}) 
compared to observed luminosity functions ({\it right, thin}) in the GC. The models assume a Salpeter IMF slope, 
an elevated lower-mass turnover of 10~\Msun, and are additionally constrained to produce 2(10$^8$)~\Msun\
in stars within the region. The observations
have been corrected for incompleteness. The third panels from the top, i.e.\ continuous star formation, best
fit the data. The observed turn-down at the faint end appears to be real and is only well fit 
only by assuming a very high lower mass turnover.\label{sfpanels}}
\end{center}
\end{figure}

One comes to similar conclusions by analyzing photometry of the field population in the GC. 
\citet{fig04b} use observed luminosity functions to determine that the star
formation rate has been roughly constant for the lifetime of the Galaxy in the GC, similar to the suggestion in \citet{ser96}
based on the sharp increase in unresolved infrared light towards the center and a mass-budget argument. Figure~\ref{sfpanels} shows model and observed luminosity
functions ({\it right}) for various star formation scenarios ({\it left}) over the lifetime of 
the Galaxy, assuming a Salpeter IMF \citep{sal55} for masses above 10~\Msun, and a flat slope below this mass. 
The observations were obtained with HST/NICMOS and have been corrected for incompleteness.
The ``burst'' models (panels 1, 2, 4, and 5) produce unrealistic ratios of bright to faint stars
in the luminosity functions, especially for the red clump near a dereddened K-band magnitude of 12.
The continuous star formation model (panel 3) best fits the data.

\section{Properties of the Three Massive Clusters}

The majority of recent star formation activity in the GC over the past 10~Myr produced three
massive clusters: the Central cluster, the Arches cluster, and the
Quintuplet cluster. The following sections describe the stellar content in
the clusters and the resultant implications for star formation in the region. They
closely follow recent reviews \citep{fig99a,fig03a,fig04}, with updates, as summarized in Table~\ref{proptable}. 

\begin{table}
\begin{center}
\begin{minipage}{12cm}
\caption{Properties of massive clusters in the Galactic Center\label{proptable}}
\symbolfootnote[0]{``M1'' is the total cluster mass in observed stars. ``M2'' is the total cluster mass in all
stars extrapolated down to a lower-mass cutoff of 1 \Msun, assuming a Salpeter IMF slope and an
upper mass cutoff of 120 \Msun (unless otherwise noted)
``Radius'' gives the average projected separation from the centroid position. 
``$\rho1$'' is M1 divided by the volume. ``$\rho2$'' is M2 divided by the volume. In either case, 
this is probably closer to the central density than
the average density because the mass is for the whole cluster while the radius is the
average projected radius. ``Age'' is the assumed age for the cluster. ``Luminosity'' gives
the total measured luminosity for observed stars. ``Q'' is the estimated Lyman continuum
flux emitted by the cluster.}
\begin{tabular}{lrrrrrrrr}
&
Log(M1) &
Log(M2) &
Radius &
Log($\rho1$) &
Log($\rho2$) &
Age &
Log(L) &
Log(Q) \\
Cluster &
\Msun &
\Msun &
pc &
\Msun \, pc$^{-3}$ &
\Msun \, pc$^{-3}$ &
\Myr &
\Lsun & 
s$^{-1}$ \\
Quintuplet & 3.0& 3.8& 1.0& 2.4& 3.2& 3$-$6& 7.5 & 50.9 \\
Arches\symbolfootnote[1]{Mass estimates have been made based upon the number of 
stars having \Minit$>$20~\Msun\ given in \citet{fig99b} and the
mass function slope in \citet{sto03}. The age, luminosity and ionizing flux are from \citet{fig02}.
}& 4.1& 4.1& 0.19 & 5.6& 5.6& 2$-$3& 8.0& 51.0 \\
Center\symbolfootnote[2]{\citet{kra95}. The mass, ``M2'' has been estimated by assuming that a total 10$^{3.5}$
stars have been formed. The age spans a range covering an initial starburst, followed by
an exponential decay in the star formation rate.}
& 3.0& 4.0& 0.23& 4.6& 5.6& 3$-$7& 7.3& 50.5 \\
\end{tabular}
\end{minipage}
\end{center}
\end{table}

The three clusters are similar in many respects, as they are all young and contain $\gtrsim$10$^4$~\Msun\ in stars. 
They have very high central stellar mass densities, up to nearly 10$^6$~\Msun~pc$^{-3}$, exceeding 
central densities in most globular clusters. They have luminosities of 10$^{7-8}$~\Lsun, and
are responsible for heating nearby molecular clouds. They also 
generate 10$^{50-51}$ ionizing photons per second, enough to account for nearby
giant HII regions. The primary difference between
the clusters is likely to be age, where the Quintuplet and Central clusters are
about twice the age of the Arches cluster. In addition, the Central cluster is unique
for its population of evolved massive stars that have broad and strong helium 
emission lines \citep[and referenes therein]{kra91}. While the Quintuplet cluster has a few similar stars \citep{geb94,fig99a}, the Central cluster
has far more as a fraction of its total young stellar population \citep{pau06}.

Table~\ref{massivetable} summarizes the massive stellar content of the clusters.

\begin{table}
\caption{Massive Stars in the Galactic Center Clusters\label{massivetable}}
\smallskip
\begin{center}
{\footnotesize
\begin{tabular}{lrrrrrrp{4cm}}
\noalign{\smallskip}
&
Age (Myr) &
O &
LBV &
WN &
WC &
RSG &
References \\
\noalign{\smallskip}
Quintuplet & 4 & 100 & 2 & 6 & 11 & 1 & \citet{fig99a,geb00,hom03} \\
Arches & 2 & 160 & 0 & $\simgr$6 & 0 & 0 & \citet{fig02} \\
Center & 4$-$7 & 100 & $\simgr$1 & $\simgr$18 & $\simgr$12 & 3 & \citet{pau06} \\
\noalign{\smallskip}
Total & & 360 & $\simgr$3 & $\simgr$29 & $\simgr$23 & 4 \\
\end{tabular}
}
\end{center}
\end{table}

\subsection{Central cluster}
\begin{figure}
\begin{center}
\includegraphics[height=9cm]{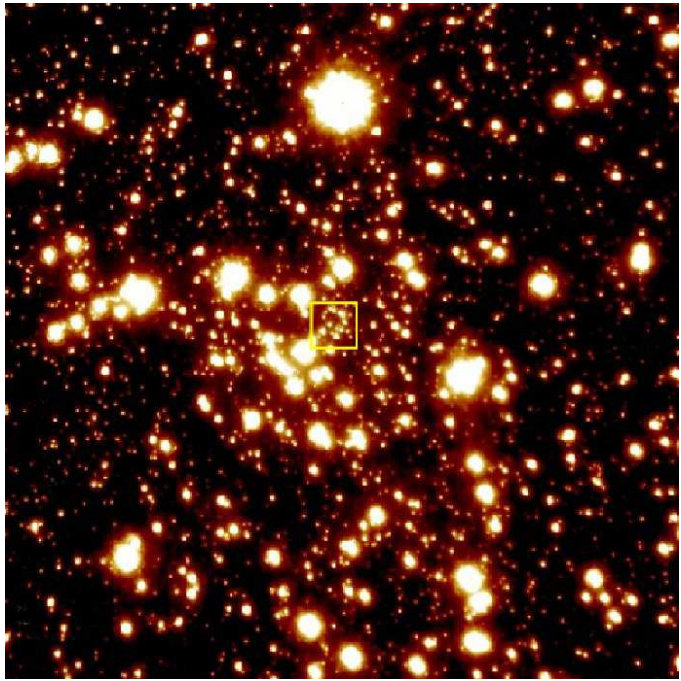}
\caption{K-band image of the Central cluster obtained with NAOS/CONICA from \citet{sch06}. The 100 or so brightest
stars in the image are evolved descendants from main sequence O-stars. The central box highlights the ``s'' stars that
are presumably young and massive (\Minit$\approx$20~\Msun). 
\label{eckartimage_annot}}
\end{center}
\end{figure}

The Central cluster contains many massive stars that have recently formed in the past 10~\Myr\
\citep{bec78,rie78,leb82,for87,all90,kra91,naj94,kra95,naj95,lib95,blu95a,blu95b,gen96,tam96,naj97}.
In all, there are now known to be at least 80 massive stars in the Central cluster \citep{eis05,pau06}, including $\approx$50 OB stars on
the main sequence and 30 more evolved massive stars (see Figure~\ref{eckartimage_annot}). 
These young stars appear to be confined to two disks \citep{gen03,lev03,pau06,tan06,bel06}.
There is also a tight collection of a dozen or so B stars (the ``s'' stars) in the central arcsecond,
highlighted in the small box in the figure.  
The formation of so many massive stars in the central parsec
remains as much a mystery now as it was at the time of the first infrared observations of the region.
Most recently, this topic has largely been supplanted by the even more improbable notion that
star formation can occur within a few thousand AU of the supermassive black hole, an idea that
will be addressed in Section~\ref{sec:sstars}.
See \citet{ale05} for a thorough review of the ``s'' stars and \citet{pau06} for a review of the
young population in the Central cluster. 

\subsection{Arches cluster}
The Arches cluster is unique in the Galaxy for its combination of extraordinarily high mass, M$\approx$10$^4$~\Msun, 
and relatively young age, $\tau=2~\Myr$\ \citep{fig02}. Being so young and massive, it contains 
the richest collection of O-stars and WNL stars in any cluster in the Galaxy \citep{cot96,ser98,fig99b,blu01,fig02}. 
It is ideally suited for testing theories that predict the shape of the IMF up to the
highest stellar masses formed (see Section~\ref{sec:imf}).

\begin{figure}
\begin{center}
\includegraphics[height=9cm]{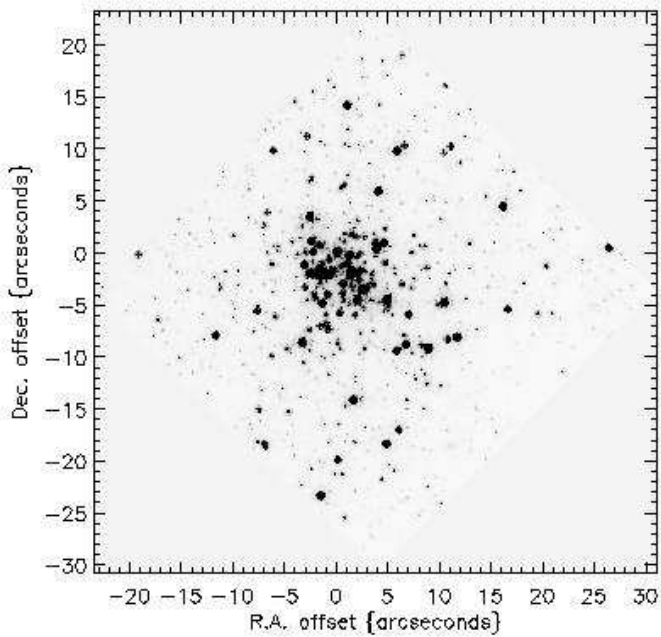}
\end{center}
\caption{F205W image of the Arches cluster obtained by \citet{fig02} using HST/NICMOS. The brightest dozen
or so stars in the cluster have \Minit$\gtrsim$100~\Msun, and there are $\approx$160 O-stars in the cluster.
The diameter is $\approx$1~lyr, making the cluster the densest in the Galaxy with $\rho>10^5~\Msun~pc^{-3}$.\label{arches}}
\end{figure}

The cluster is prominent in a broad range of observations. Figure~\ref{arches}
shows an HST/NICMOS image of the cluster -- the majority of the bright stars in the image have masses
greater than 20~\Msun. The
most massive dozen or so members of the cluster have strong emission lines at infrared
wavelengths \citep{har94,nag95,cot95,fig95t,cot96,fig99b,blu01,fig02}.
These lines are produced in strong stellar winds that are also detected at radio 
wavelengths \citep{lan01,yus03,lan05,fig02}, and x-ray 
wavelengths \citep{yus02,roc05,wan06}.

\subsection{Quintuplet cluster}
The Quintuplet cluster was originally noted for its five very bright stars, the
Quintuplet Proper Members (QPMs) \citep{gla90,oku90,nag90}. Subsequently, a number
of groups identified over 30 stars evolved from massive main sequence stars \citep{geb94,fig95,tim96,fig99a}.
Given the spectral types of the massive stars identified in the cluster, it appears that
the Quintuplet cluster is $\approx$4~\Myr\ old and had an initial mass of $>$10$^4$~\Msun\ \citep{fig99a}.
An accounting of the ionizing flux produced by the massive stars in the cluster conclusively
demonstrates that the cluster heats and ionizes the nearby ``Sickle'' HII region (see Figure~\ref{radio}).
The Quintuplet is most similar to Westerlund 1 in mass, age, and spectral content \citep{cla05,neg05,ski06,gro06,cro06}.

Of particular interest in the cluster, the QPMs are very bright at infrared wavelengths, m$_{\rm K}$ $\approx$ 6 to 9,
and have color temperatures between $\approx$ 600 to 1,000\, K. They are luminous, L$\approx$10$^{5}$~\Lsun, 
yet spectroscopically featureless, making their spectral classification ambiguous. 
\citet{fig96}, \citet{fig99a}, and \citet{mon01} argue that these objects are not protostars, OH/IR stars, or protostellar OB stars. 
Instead, they claim that these stars are dust-enshrouded WCL stars (DWCLs), 
similar to other dusty Galactic WC stars \citep{wil87}, i.e.\ WR 104 \citep{tut99} and WR 98A \citep{mon99}. 
\citet{chi03} tentatively identify a weak spectroscopic feature at 6.2~\micron\ that they
attribute to carbon, further supporting the hypothesis that these stars are indeed DWCLs. 
The stars have also been detected at x-ray wavelengths \citep{law04}, and at radio wavelengths \citep{lan99,lan05}.

Recently, \citet{tut06} convincingly show that the QPMs are indeed dusty WC stars. Figure~\ref{pinwheel} shows
data that reveal the pinwheel nature of their infrared emission, 
characteristic of binary systems containing WCL plus an OB star \citep{tut99,mon99}. This identification raises 
intruiging questions concerning massive star formation and evolution. With their identifications,
it becomes clear that every WC star in the Quintuplet is dusty, and presumably binary. 
There are two possible explanations for this result. Either the
binary fraction for massive stars is extremely high \citep{mas98,nel04}, or only binary massive stars evolve through
the WCL phase \citep{van01}.

\begin{figure}
\begin{center}
\vskip 0.3truein
\includegraphics[width=7cm]{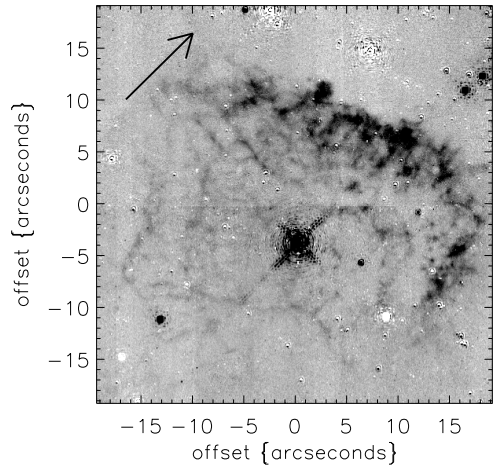}
\caption{Paschen-$\alpha$ image of the region surrounding the Pistol star from \citet{fig99c}. North is to the upper right, and east
is to the upper left. The Pistol star ejected $\approx$10~\Msun\ of material approximately 6,000~yr ago to form what
now appears to be a circumstellar nebula that is ionized by two WC stars to the north of the nebula. \citet{mon01} use ISO
data to show that the nebula is filled with dust that is heated by the Pistol star. 
\label{diffx}}
\end{center}
\end{figure}

\begin{figure}
\begin{center}
\vskip 0.3truein
\includegraphics[width=7cm]{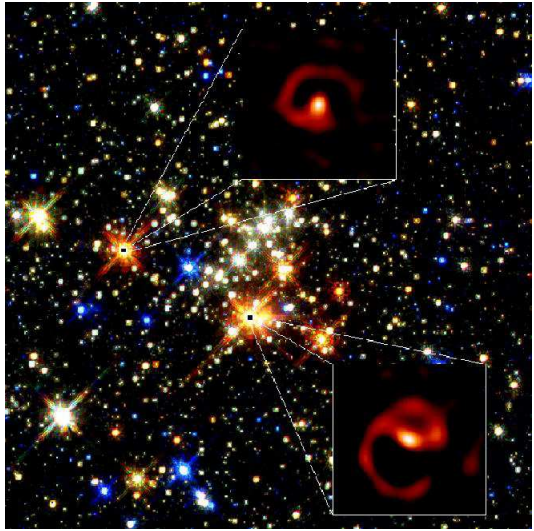}
\caption{\citet{tut06} find that the Quintuplet Proper Members are 
dusty Wolf-Rayet stars in binary systems with OB companions. The insets in this illustration show high-resolution infrared imaging 
data for two Quintuplet stars, overlaid on the HST/NICMOS image from \citet{fig99b}.
All of the Quintuplet WC stars are dusty, suggesting that they are binary.
\label{pinwheel}}
\end{center}
\end{figure}

The Quintuplet cluster also contains two Luminous Blue Variables, the Pistol star \citep{har94,fig98,fig99c},
and \FMM362 \citep{fig99a,geb00}. Both stars are extraordinarily luminous (L$>$10$^6$~\Lsun),
yet relatively cool (T$\approx$10$^4$~K), placing them in the ``forbidden zone''
of the Hertzsprung-Russell Diagram, above the Humphreys-Davidson limit \citep{hum94}. The
Pistol star is particularly intriguing, in that it is surrounded by one of the most massive (10~\Msun) circumstellar 
ejecta in the Galaxy \citep[see Figure~\ref{diffx};][]{fig99c,smi06}.
Both stars are spectroscopically \citep{fig99a} and photometrically variable \citep{gla01}.
They
present difficulties for stellar evolution and formation models. Their inferred initial masses
are $>$100~\Msun, yet such stars should have already gone supernova in a cluster that 
is so old, as evidenced by the existence of WC stars \citep{fig99a} and the red supergiant, q7 \citep{mon94,ram00}. \citet{fig02a} and \citet{fre06b} 
argue that stellar mergers might explain the youthful appearance of these stars. Alternatively, these stars might be
binary, although no evidence has been found to support this assertion. Note that in a similar case, 
LBV1806$-$20 is also surrounded by a relatively evolved cluster \citep{eik04,fig05a}, yet it
does appear to be binary \citep{fig04a}. 

\section{The initial mass function in the Galactic center\label{sec:imf}}

The IMF in the Galactic center has primarily been estimated through observations of the 
Arches cluster \citep{fig99b,sto03}, although there have been several attempts to extract such information through
observations of the Central cluster \citep{gen03,nay05,pau06} and the background population in the region \citep{fig04b}.  
These studies suggest an IMF slope that is flatter than the Salpeter value. 

\subsection{The slope\label{sec:slope}}

\citet{fig99b} and \citet{sto03} estimate a relatively flat IMF slope in the Arches cluster (see Figure~\ref{cutoff}). \citet{por02} interpret
the data to indicate an initial slope that is consistent with the Salpeter value, and 
a present-day slope that has been flattened due to dynamical evolution. Performing a similar analysis, \citet{kim00}
arrive at the opposite conclusion -- that the IMF truly was relatively flat. The primary difficulty in relating
the present-day mass function to the initial mass function is the fact that n-body interactions
operate on relatively short timescales to segregate the highest stellar masses toward the center of
the cluster and to eject the lowest stellar masses out of the cluster. Most analysis is needed to resolve
this issue.

\subsection{Upper mass cutoff\label{sec:upper}}

\begin{figure}
\begin{center}
\includegraphics[angle=90,width=10cm]{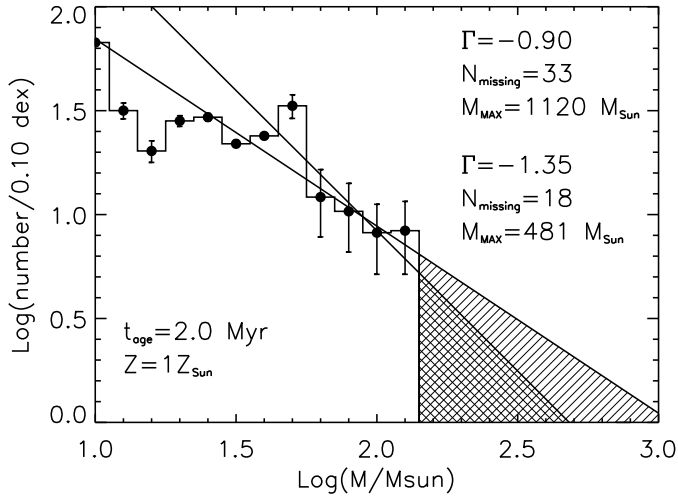}
\caption{\citet{fig05} find an apparent upper-mass cutoff to the IMF in the Arches cluster.
Magnitudes are transformed into initial mass by assuming the Geneva models for $\tau_{\rm age}$=2~\Myr,
solar metallicity, and the canonical mass-loss rates. Error bars indicate uncertainty from Poisson
statistics. Two power-law mass functions are drawn through the average of the upper four mass bins, one
having a slope of $-$0.90, as measured from the data, and another having the Salpeter slope of $-$1.35.
Both suggest a dramatic deficit of stars with \Minit$>$130~\Msun, i.e.\ 33 or 18 are missing, respectively.
These slopes would further suggest a single star with very large initial mass (M$_{\rm MAX}$). The
analysis suggests that the probability of there not being an upper-mass cutoff is $\approx$ 10$^{-8}$.
\label{cutoff}}
\end{center}
\end{figure}

The Arches cluster is the only cluster in the Galaxy that can be used to directly probe an upper
mass cutoff. It is massive enough to expect stars at least as massive as 400~\Msun, young enough for its most massive 
members to still be visible, old enough to have broken out of its natal molecular cloud, 
close enough, and at a well-established distance, for us to discern its individual stars \citep{fig05}. 
There appears to be an absence of stars with initial masses greater than 130~\Msun\ in the 
cluster, where the typical mass function predicts 18 (see Figure~\ref{cutoff}). \citet{fig05} therefore claim a firm upper mass limit 
of 150~\Msun. There is additional support for such a cutoff in other environments \citep{wei04,oey05,koe06,wei06}.

\subsection{Lower mass rollover\label{sec:lower}}

\citet{mor93} argue for an elevated lower mass rollover in the GC based on the environmental conditions therein,
and only recently have observations been deep enough to address this claim. 
\citet{sto05} claim observational evidence for an elevated cutoff around 6~\Msun\
in the Arches cluster; however, in that case, confusion and incompleteness are serious
problems. In addition, even if the apparent turn-down is a real indication of the initial cluster
population, the lack of low mass stars might result from their ejection through
n-body interactions \citep{kim00,por02}. Field observations should not suffer
from such an effect, as the field should be the repository for low mass stars ejected 
from massive clusters in the GC.
Figure~\ref{sfpanels} reveals a turn-down in the observed luminosity function of the field in the GC at a dereddened K-band magnitude greater than 16. This
appears to not be a feature of incompleteness, as the data are greater than 50\% complete at these
magnitudes \citep{fig99b}. A more convincing argument, based on this type of data, will await
even deeper observations \citep{kim06}.

\section{The ``s'' stars\label{sec:sstars}}
\begin{figure}%
\centering
\parbox{2.2in}{\includegraphics[width=2.in]{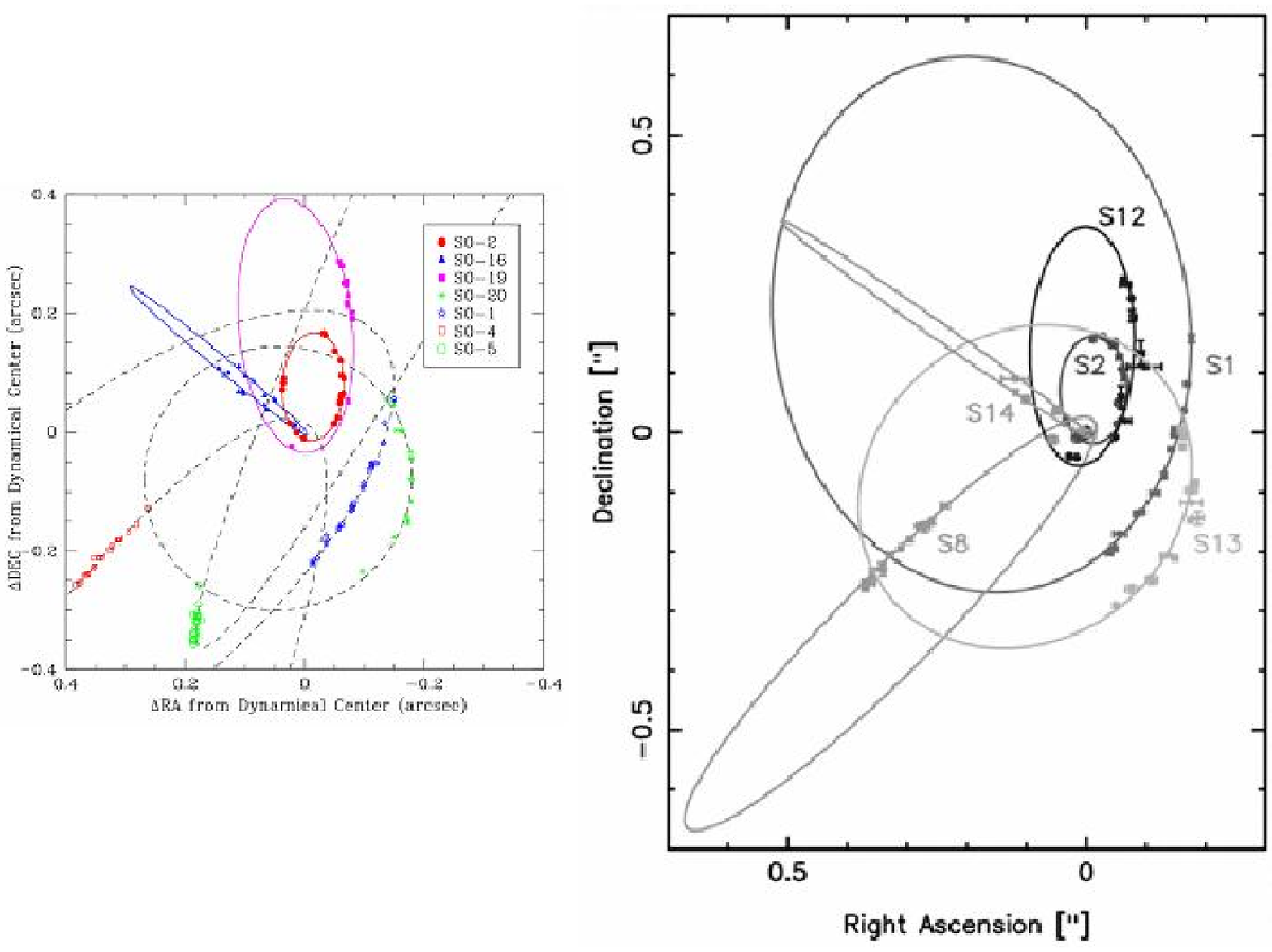}}%
\qquad
\begin{minipage}{2.2in}%
\includegraphics[width=2.in]{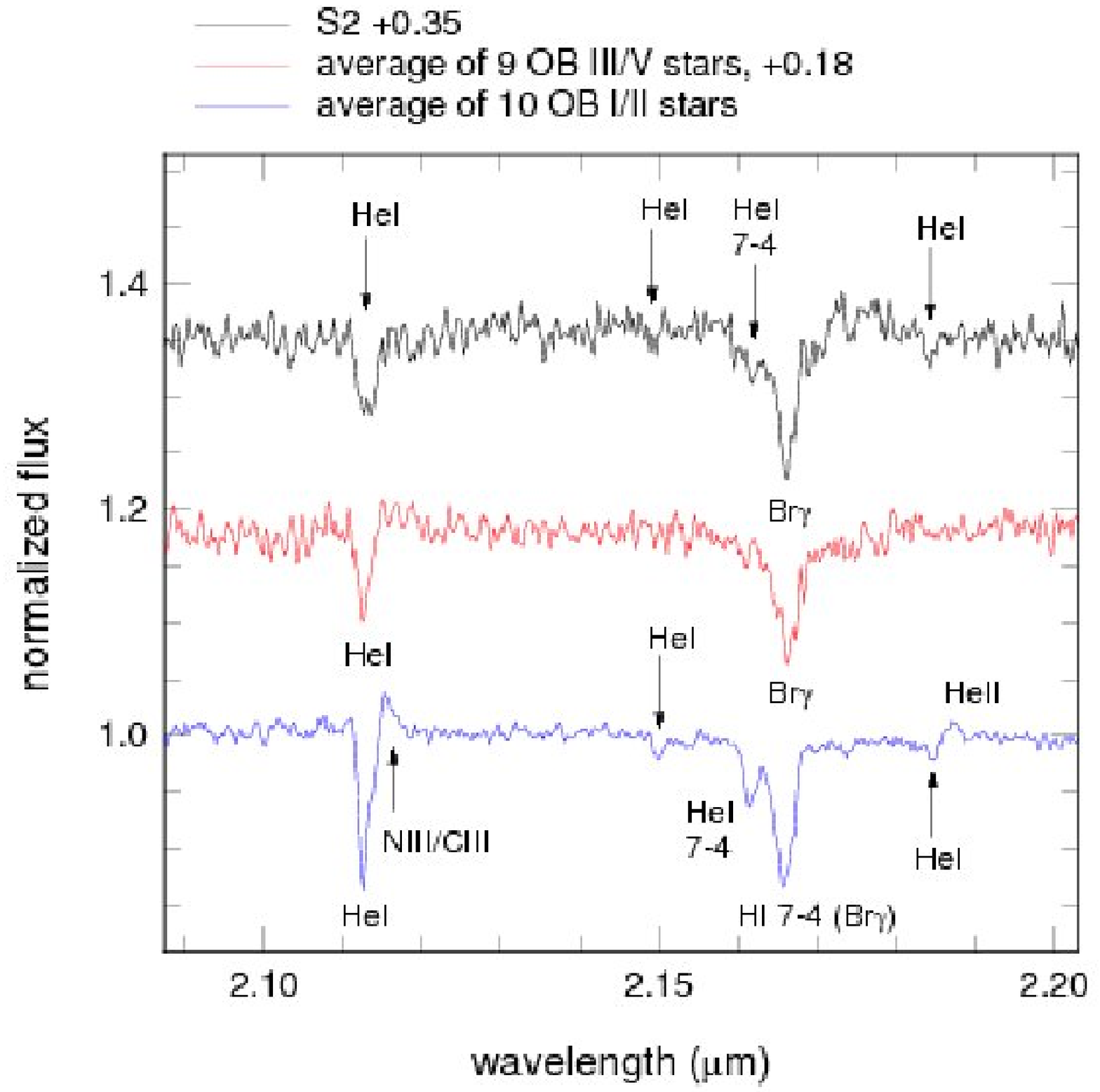}
\end{minipage}%
\caption{({\it left}) Figure~2 in \citet{ghe05} and 
({\it middle}) Figure~6 in \citet{sch03}, stretched to the same scale. Both figures fit similar model
orbits through separate proper motion data sets for the ``s'' stars. ({\it right}) \citet{pau06} find that one
of the ``s'' stars, S2, has a K-band spectrum that is similar to those of OB stars in the central parsec \citep[see also][]{ghe03}.}%
\label{combo}%
\end{figure}

Figure~\ref{eckartimage_annot} shows a dense collection of about a dozen stars within 1 arcsecond (0.04~pc) 
of Sgr~A* \citep{gen97,ghe98,ghe00,eck02,sch02,ghe03,sch03,ghe05}. 
This cluster stands out for its high stellar density, even compared to the already dense field population in the GC. 
\citet{sch03} and \citet{ghe05} (and refereces therein) have tracked 
the proper motions of the ``s'' stars, finding that they are consistent
with closed orbits surrounding a massive, and dark, object having M$\approx2-4(10^6)$~\Msun, consistent
with previous claims based on other methods \citep{lyn71,lac80,ser85,gen87,sel87,rie88,mcg89,lac91,lin92,hal96b}.
The orbital parameters for these stars are well determined, as seen in Figure~\ref{combo} ({\it left}),
and they require the existence of a supermassive black hole
in the Galactic center. While these stars are useful as gravitational test particles, they
are also interesting in their own right, as they have inferred luminosities and temperatures
that are similar to those of young 
and massive stars \citep{gen97,eck99,fig00,ghe03,eis03,eis05,pau06}.
Figure~\ref{combo} ({\it right}) shows the absorption lines that suggest relatively
high temperatures. 

\begin{table}
\begin{center}
\caption{Chronologically sorted list of references that explore hypotheses on the origin of the ``s'' stars. 
Some of the references primarily concern the other young stars in the central parsec and are included
in the table because they propose ideas that may relate to the origins of the ``s'' stars. Contributions to 
this table have been made by Tal Alexander (priv. communication).
\label{stable}}
\begin{tabular}{p{5cm}p{8cm}}
\noalign{\smallskip}
Reference &
Description \\
\noalign{\smallskip}
\citet{lac82} & tidal disruption of red giants \\
\citet{mor93} & compact objects surrounded by material from red giant envelopes (``Thorne-Zytkow objects'')\\
\citet{dav98} & red giant envelope stripping through n-body interactions \\
\citet{ale99} & red giant envelope stripping through dwarf-giant interactions   \\
\citet{bai99} & colliding red giants \\
\citet{mor99} & duty cycle of formation from infalling CND clouds and evaporation of gas reservoir by accretion and star-formation light \\
\citet{ger01} & decaying massive cluster \\
\citet{ale03} & tidal heating of stellar envelopes to form ``Squeezars'' \\
\citet{gen03} & stellar rejuvination through red giant mergers \\
\citet{gou03} & ``exchange reaction'' between massive-star binary and massive black hole \\
\citet{han03} & stars captured by inspiraling intermediate mass black hole \\
\citet{lev03} & formation in nearby gas disk \\
\citet{mcm03,kim04} & inward migration of young cluster with IMBH \\
\citet{por03,kim03} & inward migration of young cluster \\
\citet{ale04} & orbital capture of young stars by MBH-SBH binaries  \\
\citet{mil04} & formation in molecular disk \\
\citet{dav05} & tidal strippng of red giant (AGB) stars \citep[but see critique in][]{goo05} \\
\citet{gur05} & decaying cluster with formation of an IMBH \\
\citet{hai05} & formation in disk and orbital relaxation.\citep[but see critique in][]{goo05} \\
\citet{lev05} & dynamical interactions with sinking IMBH \\
\citet{nay05} & \emph{in-situ} formation within central parsec \\
\citet{nay05b} & formation in a fragmenting star disk \\
\citet{sub05} & formation in disk and accelerated orbital relaxation \\
\citet{ber06} & cluster inspiral and n-body interactions \\
\citet{hop06} & resonant relaxation of orbits \\
\citet{fre06a} & mass segregation through interactions with compact remnants \\
\citet{lev06} & star formation in fragmenting disk \\
\citet{per06} & exchange reactions between massive star binaries induced by efficient relaxation by massive perturbers \\
\citet{dra06} & tidal strippng of red giant (AGB) stars \citep[but see critique in][]{fig00}
\end{tabular}
\end{center}
\end{table}

Oddly, the increased density of the young stars in the central arcsecond is not matched
by the density distribution of old stars. Indeed, 
there is a curious absence of late-type stars in the central few arcseconds, as evidenced
by a lack of stars with strong CO absorption in their K-band spectra \citep{lac82,phi89,sel90,hal96b,gen96,gen03}.
This dearth of old stars represents a true ``hole'' in three dimensional space, and not
just a projection effect. Even the late-type stars that are projected on to the 
central parsec generally have relatively low velocities, suggesting
dynamical evidence that the region nearest to the black hole lacks old stars \citep{fig03}. 

The existence of such massive and young stars in the central arcseconds is puzzling, although
it is perhaps only an extension of the original problem in understanding the origins of the young 
stars identified in the central parsec over 20 years ago. 
Table~\ref{stable} gives a list of recent papers regarding the origin of the 
``s'' stars. While there are over 30 papers listed in this table, they can be reduced to
a few basic ideas. 
One class of ideas considers the ``s'' stars as truly young. In this case, the
``origin'' of the ``s'' stars is often reduced to the case of massive star formation in the Galactic
center region and transportation of the products to the central arcsecond. The other class
regards the ``s'' stars as old stars that only appear to be young, i.e.\ via atmospheric
stripping, merging, or heating. Both classes require new mechanisms that would be unique
to the GC, and they both have considerable weaknesses. For example, \citet{fig00} argue
that stripped red giants would not be as bright as the ``s'' stars \citep[see][for detailed
confirmation]{dra06}. See \citet{ale05} for a more thorough discussion of the strengths and
weaknesses of these ideas.

If the ``s'' stars are truly young, then that would require massive clumps to form OB stars (\Minit$\gtrsim$20~\Msun). In 
addition, the clumps would have to form from very high density material in order for them
to be stable against tidal disruption. Assuming that the stars formed as far away from
the supermassive black hole as possible, while still permitting dynamical friction to transport
them into the central arcsecond during their lifetimes, then the required densities
must be $>$10$^{11}~cm^{-3}$ \citep{fig00}. 

The average molecular cloud density in the GC
is about five orders of magnitude less, so highly compressive events might be required to 
achieve the necessary densities. Alternatively, the required densities can be reduced if 
the stars are gravitationally bound to significant mass, i.e.\ a surrounding stellar cluster. Indeed,
\citet{ger01}, \citet{por03}, and \citet{kim03} showed that particularly massive clusters could form tens of parsecs
outside of the center and be delivered into the central parsec in just a few million years.
The efficiency of this method is improved with the presence of an intermediate black hole
in the cluster \citep{mcm03,kim04}. It is key in any of these cluster transport models that
the host system have extremely high densities of $>$10$^6$~\Msun~pc$^{-3}$, comparable to 
the highest estimated central density of the Arches cluster after core collapse \citep{kim03}.
Detailed n-body simulations suggest that while these ideas may be relevant for the origins 
of the young stars in the central parsec, it is
unlikely that they could explain the existence of the ``s'' stars in the central arcsecond.



\section{Comparisons to other massive star populations in Galaxy}
There are relatively few clusters in the Galaxy with as many massive stars as in the GC clusters. NGC3603
has about a factor of two less mass than each of the GC clusters \citep{mof02}; whereas, W1
has at least a factor of two greater mass \citep{cla05,neg05,ski06,gro06}. The
next nearest similarly massive cluster is R136 in the LMC \citep{mass98}. All of these
clusters, and the GC clusters, appear to have IMF slopes that are consistent with the Salpter value (or slightly flatter) and are
young enough to still possess a significant massive star population. 
It is remarkable to note that these massive clusters appear quite similar in stellar content, whether in the Galactic disk, the
GC, or even the lower metallicity environment of the LMC. Evidently, the star formation processes,
and natal environments, that gave birth to these clusters must be similar enough to produce
clusters that are virtually indistinguishable. 

There are probably more massive clusters yet to be found in the Galaxy.
The limited sample of known massive clusters is a direct result of extinction, as
most star formation sites in the Galaxy are obscurred by dust at optical wavelengths. While infrared
observations have been available for over 30 years, they have not provided the necessary spatial
resolution, nor survey coverage, needed to probe the Galactic disk for massive 
clusters. Recently,
a number of groups have begun identifying candidate massive star clusters using near-infrared
surveys with arcsecond resolution \citep{bic03,dut03,mer05}. Indeed, these surveys have already yielded a cluster
with approximate initial mass of 20,000 to 40,000~\Msun\ \citep{fig06}, and one would expect more
to be discovered from them. 

The present-day sites of massive star formation in the Galaxy have been known for some
time through radio and far-infrared observations, as their hottest members ionize and heat nearby
gas in molecular clouds. As one of many examples, consider W49 which is the next most massive
star formation site in the Galaxy compared to Sgr~B2 \citep{hom05}, and wherein the star formation
appears to be progressing in stages over timescales that far exceed the individual collapse times
for massive star progenitors. This suggests a stimulus that triggers the star formation, perhaps
provided by a ``daisy chain'' effect in which newly formed stars trigger collapse in nearby
parts of the cloud. Similar suggestions are proposed in the 30~Dor region surrounding R136 \citep{wal02}.
While there is no evidence for an age-dispersed population in Sgr~B2, \citet{sat00} suggest that
the cloud was triggered to form stars through a cloud-cloud interaction. 

\section{Conclusions}\label{sec:concl}

Massive star formation in the GC has produced an extraordinary sample of stars populating the
initial mass function up to a cutoff of approximately 150~\Msun. The ranges of inferred masses and observed spectral
types are as expected from stellar evolution models, and the extraordinary distribution of stars in the
region is a direct consequence of the large amount of mass that has fed star formation in the GC. The origin of
the massive stars in the central parsec, and especially the central arcsecond, remains unresolved.

\begin{acknowledgments}
I thank the following individuals for discussions related to this work: Mark Morris, Bob Blum, Reinhard Genzel, 
Paco Najarro, Sungsoo Kim, and Peter Tuthill. Tal Alexander made substantial contributions to Table~3.
The material in this paper is based upon work supported by NASA
under award No.\ NNG05-GC37G, through the {\it Long Term Space Astrophysics} program.

Full resolution versions of the above images are available at 
http://www.cis.rit.edu/~dffpci/private/papers/stsci06/

\end{acknowledgments}

\end{document}